\documentclass[aps, showpacs, showkeys,nofootinbib,floatfix]{revtex4}

\usepackage{amssymb}
\usepackage{amsmath}
\usepackage{graphicx}

\begin{document}

\title{Cosmic Constraints on Holographic Dark Energy in Brans-Dicke Theory}
\author{Lixin Xu\footnote{Corresponding author}}
\email{lxxu@dlut.edu.cn}
\author{Jianbo Lu}
\author{Wenbo Li}

\affiliation{Institute of Theoretical Physics, School of Physics \&
Optoelectronic Technology, Dalian University of Technology, Dalian,
116024, P. R. China}
%
%

\begin{abstract}
In this paper, the holographic dark energy in Brans-Dicke theory is
confronted by cosmic observations from SN Ia, BAO and CMB shift
parameter. The best fit parameters are found in $1\sigma$ region:
$\Omega_{h0}=0.683^{+0.035}_{-0.038}$, $c=0.605^{+0.138}_{-0.107}$ and
$\alpha=0.00662^{+0.00477}_{-0.00467}$ (equivalently
$\omega=905.690^{+637.906}_{-651.471}$ which is less the solar system
bound and consistent with other constraint results). With these best
fit values of the parameters, it is found the universe is undergoing accelerated
expansion, and the current value of equation of state of holographic
dark energy $w_{h0}=-1.246^{+0.191}_{-0.144}$ which is phantom like in Brans-Dicke
theory. The
evolution of effective Newton's constant is also explored.
\end{abstract}

\pacs{98.80.-k, 98.80.Es, 98.80.Cq, 95.35.+d}
\keywords{Cosmology;
Holographic dark energy; Brans-Dicke theory}
\hfill TP-DUT/2009-05

\maketitle

\section{Introduction}

The observation of the Supernovae of type Ia
\cite{ref:Riess98,ref:Perlmuter99} provides the evidence that the
universe is undergoing accelerated expansion. Jointing the
observations from Cosmic Background Radiation
\cite{ref:Spergel03,ref:Spergel06} and SDSS
\cite{ref:Tegmark1,ref:Tegmark2}, one concludes that the universe at
present is dominated by $70\%$ exotic component, dubbed dark energy,
which has negative pressure and push the universe to accelerated
expansion. Of course, the accelerated expansion can attribute to the
cosmological constant naturally. However, it suffers the so-called
{\it fine tuning} and {\it cosmic coincidence} problem. To avoid
these problem, dynamic dark energy models are considered, such as
quintessence
\cite{ref:quintessence1,ref:quintessence2,ref:quintessence3,ref:quintessence4},
phtantom \cite{ref:phantom}, quintom \cite{ref:quintom} and
holographic dark energy \cite{ref:holo1,ref:holo2} etc. For recent
reviews, please see
\cite{ref:DEReview1,ref:DEReview2,ref:DEReview3,ref:DEReview4,ref:DEReview5,ref:DEReview6}.
In particular, the model, named holographic dark energy, is
constructed by considering the holographic principle and some
features of quantum gravity theory. According to the holographic
principle, the number of degrees of freedom in a bounded system
should be finite and has relations with the area of its boundary. By
applying the principle to cosmology, one can obtain the upper bound
of the entropy contained in the universe. For a system with size $L$
and UV cut-off $\Lambda$ without decaying into a black hole, it is
required that the total energy in a region of size $L$ should not
exceed the mass of a black hole of the same size, thus
$L^3\rho_{\Lambda} \le L M^2_{pl}$. The largest $L$ allowed is the
one saturating this inequality, thus $\rho_{\Lambda} =3c^2
M^{2}_{pl} L^{-2}$, where $c$ is a numerical constant and $M_{pl}$
is the reduced Planck Mass $M^{-2}_{pl}=8 \pi G$. It just means a
{\it duality} between UV cut-off and IR cut-off. The UV cut-off is
related to the vacuum energy, and IR cut-off is related to the large
scale of the universe, for example Hubble horizon, event horizon or
particle horizon as discussed by \cite{ref:holo1,ref:holo2}. In the
paper \cite{ref:holo2}, the author takes the future event horizon
\begin{equation}
R_{eh}(a)=a\int^{\infty}_{t}\frac{dt^{'}}{a(t^{'})}=a\int^{\infty}_{a}\frac{da^{'}}{Ha^{'2}}\label{eq:EH}
\end{equation}
as the IR cut-off $L$. This horizon is the boundary of the volume a
fixed observer may eventually observe. One is to formulate a theory
regarding a fixed observer within this horizon. As pointed out in
\cite{ref:holo2}, it can reveal the dynamic nature of the vacuum
energy and provide a solution to the {\it fine tuning} and {\it
cosmic coincidence} problem. In this model, the value of parameter
$c$ determines the property of holographic dark energy. When $c\ge
1$, $c=1$ and $c\le 1$, the holographic dark energy behaviors like
quintessence, cosmological constant and phantom respectively.

On the other hand, Brans-Dicke theory \cite{ref:BransDicke} as a
natural extension of Einstein's general theory of relativity can
pass the experimental tests from the solar system \cite{ref:solar}
and provide explanation to the accelerated expansion of the universe
\cite{ref:BDEp1,ref:BDEp2,ref:BDEp3}. In Brans-Dicke theory, the
gravitational constant is replaced with a inverse of time dependent
scalar field, i.e. $8\pi G=\frac{1}{\Phi(t)}$, which couples to
gravity with a coupling parameter $\omega$. The holographic dark
energy model in the framework of Brans-Dicke theory which has
already been considered by many authors
\cite{ref:BDH1,ref:BDH2,ref:BDH3,ref:BDH4,ref:BDH5,ref:BransDicke}.
In our previous paper \cite{ref:BransDicke}, the properties of
holographic dark energy in Brans-Dicke theory was discussed by
giving some characteristic values of the parameters, where the
values of the parameters were given by taking the corresponding
values obtained from constraint by cosmic observations in Einstein
theory. However, in Brans-Dicke theory, the values of parameters
would be different from that in Einstein theory. After all, the
Newton's constant $G$ evolves with time in Brans-Dicke theory. So,
there would be some differences. In fact, in our previous paper
\cite{ref:BransDicke}, the value of $\left|\omega\right|
>40000$ is taken for granted. It would be dangerous because of
the possibility of small value in large scale, say in cosmological
scale, reported by the authors \cite{ref:smallomega}. So, the
holographic dark energy in Brans-Dicke theory must be tested by
cosmic observations. This is the main task of this work. In this
paper, the cosmic observations from SN Ia, BAO and CMB shift
parameter will be used as cosmic constraints, for details please see
the following sections. When using these observational data set, one
has to notice the evolution of Newton's constant $G$. In fact, the
SN Ia as a useful cosmic constraint has been considered in
\cite{ref:smallomega,ref:BDSN1,ref:BDSN2,ref:BDSN3}.

This paper is structured as follows. In Section \ref{sec:HDBD}, we
give a brief review of holographic dark energy in Brans-Dicke
theory. A brief description of comic observations and methods used
in our paper is listed in Section \ref{sec:constraints}. Section
\ref{sec:results} is the results and discussion.

\section{Holographic dark energy in Brans-Dicke theory}\label{sec:HDBD}

Here, we just give a brief review of holographic dark energy in
Brans-Dicke theory, for the details please see \cite{ref:BDHD}. The
holographic dark energy in Brans-Dicke theory takes the form
\begin{equation}
\rho_{h}=3c^2 \Phi(t)L^{-2},
\end{equation}
where $\Phi(t)=\frac{1}{8 \pi G_{eff}}$ is a reverse of time
variable Newton's constant. In a spatially flat FRW cosmology filled with
dark matter and holographic dark energy, the gravitational equations
can be written as
\begin{eqnarray}
3\Phi\left[H^2+H\frac{\dot{\Phi}}{\Phi}-\frac{\omega}{6}\frac{\dot{\Phi}^2}{\Phi^2}\right]=\rho_{m}+\rho_{h},\label{eq:FE}\\
2\frac{\ddot{a}}{a}+H^2+\frac{\omega}{2}\frac{\dot{\Phi}^2}{\Phi^2}+2H\frac{\dot{\Phi}}{\Phi}+\frac{\ddot{\Phi}}{\Phi}=-\frac{p_{h}}{\Phi},\label{eq:RE}
\end{eqnarray}
where $H=\frac{\dot{a}}{a}$ is the Hubble parameter, $\rho_{m}$ is
dark matter energy density, $\rho_{h}$ is the holographic dark
energy density and $p_{h}$ is the pressure of holographic dark
energy. The scalar field evolution equation is
\begin{equation}
\ddot{\Phi}+3H\dot{\Phi}=\frac{\rho_{m}+\rho_{h}-3p_{h}}{2\omega+3}.\label{eq:phiE}
\end{equation}
Considering the dark matter energy conservation equation
\begin{equation}
\dot{\rho_{m}}+3H\rho_{m}=0,
\end{equation}
and jointing it with Eq. (\ref{eq:FE}), Eq. (\ref{eq:RE}) and Eq.
(\ref{eq:phiE}), one obtains the holographic dark energy
conservation equation
\begin{equation}
\dot{\rho_{h}}+3H(\rho_{h}+p_{h})=0.\label{eq:DECE}
\end{equation}
Here, we have considered non-interacting cases. The Friedmann
equation (\ref{eq:FE}) is
\begin{equation}
H^2=\frac{\rho_{m}+\rho_{h}}{3\Phi}-H\frac{\dot{\Phi}}{\Phi}+\frac{\omega}{6}\frac{\dot{\Phi}^2}{\Phi^2}.
\label{eq:SFE}
\end{equation}
With the assumption $\Phi/\Phi_0=(a/a_0)^{\alpha}$, the Eq.
(\ref{eq:SFE}) is rewritten as
\begin{equation}
H^2=\frac{2}{(6+6\alpha-\omega\alpha^2)\Phi}(\rho_{m}+\rho_{h}).\label{eq:RFE}
\end{equation}
It is easy to find out that, in the limit case $\alpha \rightarrow
0$, the standard cosmology is recovered. To make the Friedmann
equation (\ref{eq:RFE}) to have physical meanings, i.e. to make
$(6+6\alpha-\omega\alpha^2)>0$, one has the following constraints on
the values of $\alpha$
\begin{equation}
\begin{array}{ccc}
\frac{3-\sqrt{9+6\omega}}{\omega}<\alpha<\frac{3+\sqrt{9+6\omega}}{\omega},
& \text{for} & \omega>0,\\
\alpha<\frac{3-\sqrt{9+6\omega}}{\omega} \quad \text{or} \quad
\alpha>\frac{3+\sqrt{9+6\omega}}{\omega},& \text{for} & -3/2
\leq \omega<0,\\
\Re, & \text{for} & \omega < -3/2.
\end{array}\label{eq:condition}
\end{equation}
However, the solar system experiments predict the value of $\omega$
is $\left|\omega\right| > 40000$ \cite{ref:solar}. However, the
value of parameter $\omega =-3/2$ is a boundary of ghost
\cite{ref:BDghost}. So, in this paper, when considering these
constraints, the second line of Eq. (\ref{eq:condition}) will be
omitted and $\omega> 0$ will be consider in this paper. In fact,
authors \cite{ref:BDCos} have used the cosmic observations to
constrain the parameter $\omega$. In \cite{ref:BDCos}, the authors
found that $\omega$ can be smaller than $40000$ in cosmological
scale, say $\omega\sim1000$.

When the event horizon $R_{eh}$
\begin{equation}
R_{eh}(a)=a\int^{\infty}_{t}\frac{dt^{'}}{a(t^{'})}=a\int^{\infty}_{a}\frac{da^{'}}{Ha^{'2}}\label{eq:EH}
\end{equation}
is taken as the IR cut-off. The holographic dark energy is
\begin{equation}
\rho_{h}=\frac{3c^2\Phi}{R^2_{eh}}.\label{eq:EHHDE}
\end{equation}
And, the Friedmann Eq. (\ref{eq:RFE}) is rewritten as
\begin{eqnarray}
H^2&=&H^2_0\Omega_{m0}\left(\frac{a_0}{a}\right)^{(3+\alpha)}+\Omega_{h}H^2
\nonumber \\
&=&H^2_0\Omega_{m0}a^{-(3+\alpha)}+\Omega_{h}H^2,\label{eq:EHFE}
\end{eqnarray}
where the dimensionless energy density of of dark matter is
$\Omega_{m0}=\frac{2}{(6+6\alpha-\omega
\alpha^2)}\frac{\rho_{m0}}{\Phi_0 H^2_0}\equiv\frac{8\pi
G\rho_{m0}}{3H^2_0}$ and the one of holographic dark energy
$\Omega_{h}$ is the solution of differential equation
\begin{equation}
\Omega_{h}'=\Omega_{h}\left(1-\Omega_{h}\right)\left(1+\alpha+\frac{2}{c}\sqrt{\Omega_{h}}\right),\label{eq:diffeq}
\end{equation}
where $'$ denotes the derivative with respect to $x =\ln a$. This
equation describes the evolution of dimensionless energy density of
dark energy. Comparing the definition of $\Omega_{m0}$ with that of the standard cosmological model, one easily has the relation $\omega
\alpha=6$. With the relation $a_0/a=1+z$, the Friedmann equation
(\ref{eq:EHFE}) can be rewritten as
\begin{equation}
H^2=H^2_0\frac{\Omega_{m0}a^{-(3+\alpha)}}{1-\Omega_{h}}=H^2_0\frac{\Omega_{m0}(1+z)^{(3+\alpha)}}{1-\Omega_{h}},
\end{equation}
From the conservation equation of the holographic dark energy
(\ref{eq:DECE}), on has the equation of state (EoS) of holographic
dark energy
\begin{equation}
w_{h}=-1-\frac{1}{3}\frac{d \ln \rho_{h}}{d \ln
a}=-\frac{1}{3}\left(1+\alpha+\frac{2}{c}\sqrt{\Omega_{h}}\right),\label{eq:DEEOS}
\end{equation}
where $w_{h}=p_{h}/\rho_{h}$. From the above equation, one finds the
EoS of holographic dark energy is in the range of
\begin{equation}
-\frac{1}{3}\left(1+\alpha+\frac{2}{c}\right)<w_{h}<-\frac{1}{3}\left(1+\alpha\right),
\end{equation}
when one considers the holographic dark energy density ratio
$0\leq\Omega_{h}\leq1$.
Also, by using the Eq. (\ref{eq:RE}) and the assumption
$\Phi/\Phi_0=(a/a_0)^{\alpha}$, one obtains the deceleration
parameter as follows
\begin{equation}
q=-\frac{\ddot{a}}{aH^2}=\frac{1}{2}+\alpha+\frac{\alpha}{8+2\alpha}+\frac{6w_h\Omega_{h}}{4+\alpha}.\label{eq:dec}
\end{equation}
It is clear that the 'Standard' holographic dark energy will be
recovered in the limit $\alpha \rightarrow 0$. In Brans-Dicke theory
case of holographic dark energy, the properties of the holographic
dark energy are determined by the best fit values of parameters $c$ and $\alpha$ which
would be obtained by confronting with cosmic observational data. It
can be easily seen that the holographic dark energy can be
quintessence, phantom and quitom as that in the Standard case. But,
all these properties must be determined by cosmic observations.

\section{Cosmic Observational Constraints}\label{sec:constraints}

In this section, cosmic observations and methods used in this paper
are described.

\subsection{SN Ia}

We constrain the parameters with the Supernovae Cosmology Project
(SCP) Union sample including $307$ SN Ia \cite{ref:SCP}, which
distributed over the redshift interval $0.015\le z\le 1.551$.
Constraints from SN Ia can be obtained by fitting the distance
modulus $\mu(z)$ \cite{ref:smallomega,ref:BDSN1,ref:BDSN2,ref:BDSN3}
\begin{equation}
\mu_{th}(z)=5\log_{10}(D_{L}(z))+\frac{15}{4}\log_{10}\frac{G_{eff}}{G}+\mu_{0},
\end{equation}
where, $G$ is the current value of effective Newton's constant
$G_{eff}$, $D_{L}(z)$ is the Hubble free luminosity distance $H_0
d_L(z)/c$ and
\begin{eqnarray}
d_L(z)&=&c(1+z)\int_{0}^{z}\sqrt{\frac{G}{G(z')}}\frac{dz^{\prime}}{H(z^{\prime})}\\
\mu_0&\equiv&42.38-5\log_{10}h,
\end{eqnarray}
where $H_0$ is the Hubble constant which is denoted in a
re-normalized quantity $h$ defined as $H_0 =100 h~{\rm km ~s}^{-1}
{\rm Mpc}^{-1}$. The observed distance moduli $\mu_{obs}(z_i)$ of SN
Ia at $z_i$ is
\begin{equation}
\mu_{obs}(z_i) = m_{obs}(z_i)-M,
\end{equation}
where $M$ is their absolute magnitudes.

For SN Ia dataset, the best fit values of parameters in a model can
be determined by the likelihood analysis is based on the calculation
of
\begin{eqnarray}
\chi^2(p_s,m_0)&\equiv& \sum_{SNIa}\frac{\left[
\mu_{obs}(z_i)-\mu_{th}(p_s,z_i)\right]^2} {\sigma_i^2} \nonumber\\
&=&\sum_{SNIa}\frac{\left[ 5 \log_{10}(D_L(p_s,z_i)) - m_{obs}(z_i)
+ m_0 \right]^2} {\sigma_i^2}, \label{chi2}
\end{eqnarray}
where $m_0\equiv\mu_0+M$ is a nuisance parameter (containing the
absolute magnitude and $H_0$) that we analytically marginalize over
\cite{ref:SNchi2},
\begin{equation}
\tilde{\chi}^2(p_s) = -2 \ln \int_{-\infty}^{+\infty}\exp \left[
-\frac{1}{2} \chi^2(p_s,m_0) \right] dm_0 \; ,
\label{chi2_marginalization}
\end{equation}
to obtain
\begin{equation}
\tilde{\chi}^2 =  A - \frac{B^2}{C} + \ln \left(
\frac{C}{2\pi}\right) , \label{chi2_marginalized}
\end{equation}
where
\begin{equation}
A=\sum_{SNIa} \frac {\left[5\log_{10}
(D_L(p_s,z_i))-m_{obs}(z_i)\right]^2}{\sigma_i^2},
\end{equation}
\begin{equation}
B=\sum_{SNIa} \frac {5
\log_{10}(D_L(p_s,z_i)-m_{obs}(z_i)}{\sigma_i^2},
\end{equation}
\begin{equation}
C=\sum_{SNIa} \frac {1}{\sigma_i^2} \; .
\end{equation}
The Eq. (\ref{chi2}) has a minimum at the nuisance parameter value
$m_0=B/C$. Sometimes, the expression
\begin{equation}
\chi^2_{SNIa}(p_s,B/C)=A-(B^2/C)\label{eq:chi2SNIa}
\end{equation}
is used instead of Eq. (\ref{chi2_marginalized}) to perform the
likelihood analysis. They are equivalent, when the prior for $m_0$
is flat, as is implied in (\ref{chi2_marginalization}), and the
errors $\sigma_i$ are model independent, what also is the case here.
Obviously, from the value $m_0=B/C$, one can obtain the best-fit
value of $h$ when $M$ is known.

To determine the best fit values of parameters for each model, we minimize
$\chi^2(p_s,B/C)$ which is equivalent to maximizing the likelihood
\begin{equation}
{\cal{L}}(p_s) \propto e^{-\chi^2(p_s,B/C)/2} .
\end{equation}

\subsection{BAO}
The BAO are detected in the clustering of the combined 2dFGRS and
SDSS main galaxy samples, and measure the distance-redshift relation
at $z = 0.2$. BAO in the clustering of the SDSS luminous red
galaxies measure the distance-redshift relation at $z = 0.35$. The
observed scale of the BAO calculated from these samples and from the
combined sample are jointly analyzed using estimates of the
correlated errors, to constrain the form of the distance measure
$D_V(z)$ \cite{ref:Okumura2007,ref:Eisenstein05,ref:Percival}
\begin{equation}
D_V(z)=\left[(1+z)^2 D^2_A(z) \frac{cz}{H(z)}\right]^{1/3},
\label{eq:DV}
\end{equation}
where $D_A(z)$ is the proper (not comoving) angular diameter
distance which has the following relation with $d_{L}(z)$
\begin{equation}
D_A(z)=\frac{d_{L}(z)}{(1+z)^2}.
\end{equation}
Matching the BAO to have the same measured scale at all redshifts
then gives \cite{ref:Percival}
\begin{equation}
D_{V}(0.35)/D_{V}(0.2)=1.812\pm0.060.
\end{equation}
Then, the $\chi^2_{BAO}(p_s)$ is given as
\begin{equation}
\chi^2_{BAO}(p_s)=\frac{\left[D_{V}(0.35)/D_{V}(0.2)-1.812\right]^2}{0.060^2}\label{eq:chi2BAO}.
\end{equation}

\subsection{CMB shift Parameter R}

The CMB shift parameter $R$ is given by \cite{ref:Bond1997}
\begin{equation}
R(z_{\ast})=\sqrt{\Omega_m H^2_0}(1+z_{\ast})D_A(z_{\ast})/c
\end{equation}
which is related to the second distance ratio
$D_A(z_\ast)H(z_\ast)/c$ by a factor $\sqrt{1+z_{\ast}}$. Here the
redshift $z_{\ast}$ (the decoupling epoch of photons) is obtained by
using the fitting function \cite{Hu:1995uz}
\begin{equation}
z_{\ast}=1048\left[1+0.00124(\Omega_bh^2)^{-0.738}\right]\left[1+g_1(\Omega_m
h^2)^{g_2}\right],
\end{equation}
where the functions $g_1$ and $g_2$ are given as
\begin{eqnarray}
g_1&=&0.0783(\Omega_bh^2)^{-0.238}\left(1+ 39.5(\Omega_bh^2)^{0.763}\right)^{-1}, \\
g_2&=&0.560\left(1+ 21.1(\Omega_bh^2)^{1.81}\right)^{-1}.
\end{eqnarray}
The 5-year {\it WMAP} data of $R(z_{\ast})=1.710\pm0.019$
\cite{ref:Komatsu2008} will be used as constraint from CMB, then the
$\chi^2_{CMB}(p_s)$ is given as
\begin{equation}
\chi^2_{CMB}(p_s)=\frac{(R(z_{\ast})-1.710)^2}{0.019^2}\label{eq:chi2CMB}.
\end{equation}

\section{Results and Discussion}\label{sec:results}

\begin{figure}[tbh]
\centering
\includegraphics[width=2.9in]{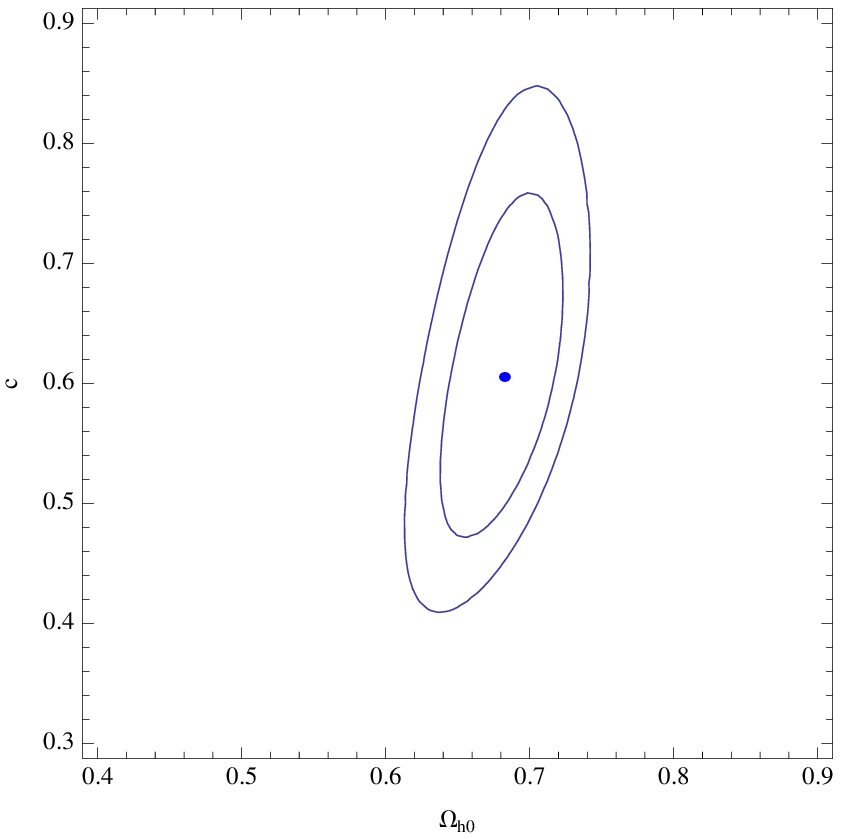}
\includegraphics[width=3.in]{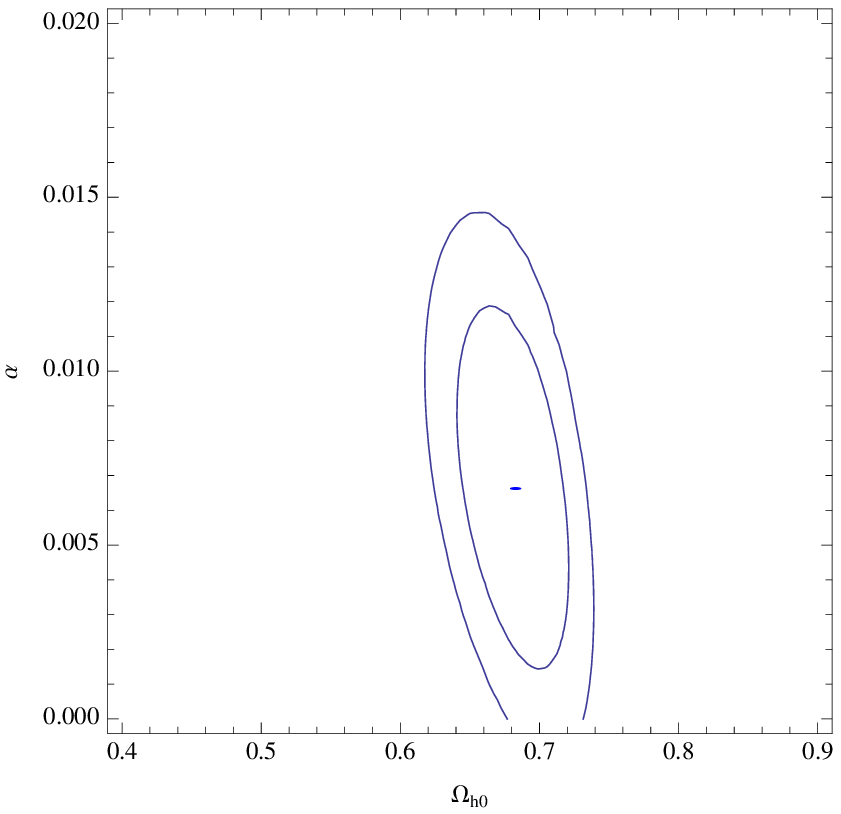}
\caption{The contour plot of $\Omega_{h0}-c$ and
$\Omega_{h0}-\alpha$ with $1\sigma,2\sigma$ confidence
levels.}\label{fig:contour}
\end{figure}

For Gaussian distributed measurements, the likelihood function
$L\propto e^{-\chi^2/2}$, where $\chi^2$ is
\begin{equation}
\chi^2=\chi^2_{SNIa}+\chi^2_{BAO}+\chi^2_{CMB},
\end{equation}
where $\chi^2_{SNIa}$ is given in Eq. (\ref{eq:chi2SNIa}),
$\chi^2_{BAO}$ is given in Eq. (\ref{eq:chi2BAO}), $\chi^2_{CMB}$ is
given in Eq. (\ref{eq:chi2CMB}). In this paper, the central values
of $\Omega_b h^2=0.02265\pm 0.00059$, $\Omega_m h^2=0.1369\pm
0.0037$ from 5-year {\it WMAP} results \cite{ref:Komatsu2008} and
$H_0=72\pm{\rm 8 km s^{-1}Mpc^{-1}}$ are adopted. After calculation,
the results are listed in Tab. \ref{tab:result}.
\begin{table}[tbh]
\begin{center}
\begin{tabular}{c|c|c|c|c|c}
\hline\hline Datasets & $\chi^2_{min}$ & $\Omega_{h0}(1\sigma)$ & $c(1\sigma)$ & $\alpha(1\sigma)$ & $\omega(1\sigma)$\\
\hline SN+BAO+CMB & $317.156$  & $0.683^{+0.035}_{-0.038}$ & $0.605^{+0.138}_{-0.107}$ & $0.00662^{+0.00477}_{-0.00467}$ & $905.690^{+637.906}_{-651.471}$ \\
\hline
\end{tabular}
\caption{The minimum values of $\chi^2$ and best fit values of the
parameters.}\label{tab:result}
\end{center}
\end{table}
The contours of $\Omega_{h0}-c$ and $\Omega_{h0}-c$ with $1\sigma,2\sigma$ confidence
levels are plotted in Fig. \ref{fig:contour}.

Current constraints \cite{ref:VG} on the variation of Newton's
constant imply
\begin{equation}
\left|\frac{\dot{G}_{eff}}{G_{eff}}\right|\leq 10^{-11}yr^{-1},
\end{equation}
in our case, which corresponds to
\begin{equation}
\left|\frac{\dot{\Phi}}{\Phi}\right|=\alpha H \leq 10^{-11}yr^{-1}.
\end{equation}
It implies
\begin{equation}
\alpha \leq \frac{1}{H}\times 10^{-11}yr^{-1}.
\end{equation}
Considering the current value of Hubble constant
$h=0.72^{+0.08}_{-0.08}$, one obtains the bounds on $\alpha$, when
the central value is taken
\begin{equation}
\alpha \leq 0.135807.
\end{equation}
It is clear that the best fit value of parameter $\alpha$ is under the bound and consistent.

\begin{figure}[tbh]
\centering
\includegraphics[width=4.5in]{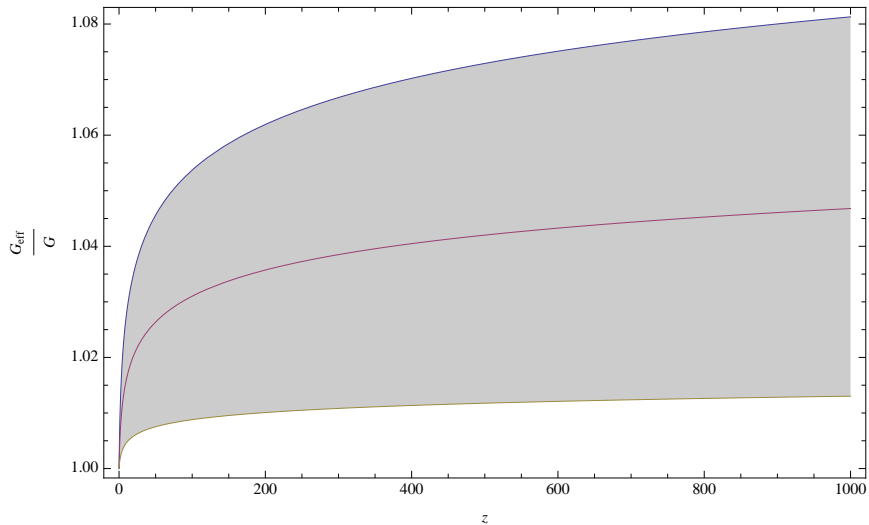}
\caption{The evolution of effective Newton's constant with
redshift $$z, with $1\sigma$ regions.}\label{fig:G}
\end{figure}

The evolution of the effective Newton's constant is written as follows
\begin{equation}
\frac{G_{eff}}{G}=(1+z)^\alpha
\end{equation}
under the assumption $\Phi/\Phi_{0}=(a/a_0)^{\alpha}$.
With the best fit value of the parameter $\alpha=0.00662$, the evolution of effective Newton's constant with redshift $z$ is
plotted in Fig. \ref{fig:G} with the best fit parameter.

\section{Conclusions}\label{sec:conclusion}

In this paper, the holographic dark energy in Brans-Dicke theory is
constrained by cosmic observations which include the data sets from
SN Ia, BAO and CMB shift parameter. In $1\sigma$ region, the best
values of the parameters are: $\Omega_{h0}=0.683^{+0.035}_{-0.038}$, $c=0.605^{+0.138}_{-0.107}$ and
$\alpha=0.00662^{+0.00477}_{-0.00467}$. Equivalently, with the relation $\omega\alpha=6$ holds, the best fit
value of $\omega$ in $1\sigma$ region is $\omega=905.690^{+637.906}_{-651.471}$ which is smaller than the value from the solar system bound, but
consistent with other reports in cosmological scale
\cite{ref:smallomega}. With these best fit values of parameters, it
is found the universe is undergoing accelerated expansion currently,
and the current value of equation of state of holographic dark
energy $w_{h0}=-1.246^{+0.191}_{-0.144}$ which is phantom like in Brans-Dicke theory. The
evolution of effective Newton's constant is explored. Its best fit
value is consistent with the bound \cite{ref:VG}.

\acknowledgements{This work is supported by NSF (10703001), SRFDP
(20070141034) of P.R. China.}

\end{document}